\documentclass[preprint,aps,nofootinbib]{revtex4-1}

\usepackage{graphicx}
\usepackage{amsmath,slashed}
\usepackage{amsfonts}
\usepackage{hyperref}
\usepackage[capitalise]{cleveref}
\usepackage{epstopdf}
\usepackage{subcaption}
\usepackage{multirow}
\usepackage{fancyhdr}
\usepackage[utf8]{inputenc}
\usepackage{xcolor}

\def\be{\begin{equation}}
\def\ee{\end{equation}}

\def\vep{\varepsilon}

\def\tr{\mathrm{tr}}
\def\ordr{\mathcal{O}}

\def\sib{\begin{align}r{\sigma}}
\def\<{\langle}
\def\>{\rangle}

\def\mt{{\mathcal{T}}}

\def\lag{{\mathcal{L}}}

\def\non{\nonumber}
\def\bea{\begin{eqnarray}}
\def\eea{\end{eqnarray}}
\def\bean{\begin{eqnarray*}}
\def\eean{\end{eqnarray*}}

\def\nlsm{\text{NLSM}}

\def\ee{\eta}

\def\ta{{\tilde{a}}}
\def\tb{{\tilde{b}}}
\def\tc{{\tilde{c}}}

\def\SO{{\text{SO}}}

\def\U{{\text{U}}}

\def\pf{\mathrm{Pf}}
\def\A{\textsf{A}}

\def\ta{{\tilde{a}}}
\def\tb{{\tilde{b}}}
\def\tc{{\tilde{c}}}

\def\tdf{{\tilde{f}}}

\def\uphi{\underline{\phi}}
\def\pt{\mathcal{C}}
\def\ss{{\text{ss}}}
\def\nl{{\text{nl}}}

\def\hl{{\text{hl}}}

\def\sj{\textsf{j}}
\def\sJ{\textsf{J}}
\def\sd{\textsf{d}}
\def\sff{\textsf{f}}
\def\gT{\textsf{T}}
\def\gX{\textsf{X}}
\def\sX{{ X}}
\def\sY{{ Y}}
\def\M{{\cal M}}
\def\edbi{\text{eDBI}}
\def\Pf{\text{Pf}}
\def\cin{{\cal I}}

\def\uf{\underline{F}}

\newcommand{\Rmnum}[1]{\expandafter\@slowromancap\romannumeral #1@}

\begin{document}

\vspace*{.5cm}

\title{Double Copy in Higher Derivative Operators of Nambu-Goldstone Bosons}

\author{\vspace{0.5cm} Ian Low$^{\, a,b}$, Laurentiu Rodina$\, ^c$, Zhewei Yin$^{\, b}$}
\affiliation{\vspace{0.5cm}
\mbox{$^a$ High Energy Physics Division, Argonne National Laboratory, Lemont, IL 60439, USA}\\
\mbox{$^b$ Department of Physics and Astronomy, Northwestern University, Evanston, IL 60208, USA}\\ 
\mbox{$^c$ Institut de Physique Theorique, Universite Paris Saclay,}\\
\mbox{CEA, CNRS, F-91191 Gif-sur-Yvette, France}
 \vspace{0.5cm}
}

\begin{abstract}
We investigate the existence of double copy structure, or the lack thereof, in higher derivative operators for Nambu-Goldstone bosons. At the leading ${\cal O}(p^2)$, tree amplitudes of Nambu-Goldstone bosons in the adjoint representation can be (trivially) expressed as the double copy of itself and the cubic bi-adjoint scalar theory, through the Kawai-Lewellen-Tye bilinear kernel.  At the next-to-leading ${\cal O}(p^4)$ there exist four operators in general, among which we identify one  operator whose amplitudes exhibit the flavor-kinematics duality and can be written as the double copy of ${\cal O}(p^2)$ Nambu-Goldstone amplitudes and the Yang-Mills+$\phi^3$ theory, involving both gluons and gauged cubic bi-adjoint scalars. The specific operator turns out to coincide with the scalar ${\cal O}(p^4)$ operator in the so-called extended Dirac-Born-Infeld theory, for which the aforementioned double copy relation holds more generally.

\end{abstract}

\maketitle

\tableofcontents

\section{Introduction}
The nonlinear sigma model (NLSM) \cite{GellMann:1960np,Coleman:1969sm,Callan:1969sn} is an effective field theory (EFT) of Nambu-Goldstone bosons (NGB's) arising from spontaneously broken symmetries.  Recent developments in the modern S-matrix program have led to renewed interest in the NLSM, which is frequently referenced. In particular, NLSM can be formulated in an entirely on-shell way, by imposing the consistency condition of the Adler's zero \cite{Adler:1964um,Susskind:1970gf}. This powerful on-shell property,  related to a shift symmetry at the Lagrangian level \cite{Low:2014nga,Low:2015ogb,Low:2017mlh,Low:2018acv},  leads to a wealth of constructions from totally different perspectives, including soft bootstrap \cite{Cheung:2015ota,Elvang:2018dco,Low:2019ynd} and single soft scaling or double soft theorems \cite{Kampf:2013vha,Arkani-Hamed:2016rak,Rodina:2016jyz,Rodina:2018pcb}.

Furthermore, the NLSM is also a key element of the color-kinematics duality and the ensuing Bern-Carrasco-Johansson (BCJ) double copy \cite{Bern:2008qj}, as well as the Cachazo-He-Yuan (CHY) formalism for S-matrix \cite{Cachazo:2013gna,Cachazo:2013hca,Cachazo:2013iea,Cachazo:2014xea}. These formalisms have demonstrated a remarkable unity among naively distinct theories, by expressing for instance gravity as the double copy of Yang-Mills (YM), or Born-Infeld as the double copy between YM and NLSM \cite{Cachazo:2014xea}. At the leading ${\cal O}(p^2)$, the requirement of flavor-kinematics duality, together with locality and cyclic invariance of flavor-ordered amplitudes, can even uniquely constrain tree amplitudes in the NLSM  \cite{Carrasco:2019qwr}.

More generally, in the space of consistent quantum  theories, the NLSM can be related to YM through transmutation operators and dimensional reduction \cite{Cheung:2016prv,Cheung:2017ems,Cheung:2017yef} or to bi-adjoint scalar through soft limits \cite{Cachazo:2016njl,Low:2018acv,Mizera:2018jbh,Yin:2018hht}. These fascinating aspects are somewhat hidden in the traditional Lagrangian formulation. Finally, through a subset of higher derivative corrections starting from ${\cal O}(p^6)$, it also makes an appearance in string theory, as symmetrized sums over Z-theory amplitudes, which are objects carrying the $\alpha^\prime$ dependence of the superstring theory \cite{Carrasco:2016ldy,Carrasco:2016ygv}. The origin in string theory ensures the BCJ relation derived at ${\cal O}(p^2)$ is satisfied by all higher derivative operators in the Z-theory.

From the effective field theory perspective, a natural puzzle arises when one includes generic higher dimensional and higher derivative corrections to the leading renormalizable interactions: do any of the fascinating features, such as the double copy relation, continue to work in these cases? Some preliminary studies showed that for NLSM, direct applications of the BCJ relations from ${\cal O}(p^2)$ fail at ${\cal O}(p^4)$ \cite{Elvang:2018dco,Carrillo-Gonzalez:2019aao}.

However, recently new ingredients for constructing the color-kinematics duality are introduced at the level of 4-pt amplitudes \cite{Carrasco:2019yyn,Low:2019wuv,Carrasco5point}, which involve new color (flavor) kinematic objects as linear combinations of color (flavor) structures with coefficients given by Mandelstam invariants. In this paper we extend the results on color-kinematics duality beyond the 4-pt amplitudes, to higher multiplicity, and investigate whether it is possible to construct double copy relations for NLSM at ${\cal O}(p^4)$.

It is also instructive to consider the double copy relation from the Kawai-Lewellen-Tye (KLT) bilinear form, which for NLSM at the leading ${\cal O}(p^2)$  can be expressed as \cite{Chiodaroli:2014xia,Cachazo:2014xea}:
\[
 \nlsm^{(2)} =  \nlsm^{(2)} \stackrel{\rm KLT}{\otimes} \phi^3  \ , 
\]
where $\phi^3$ denotes the cubic bi-adjoint scalar theory, and the universal KLT kernel matrix $\stackrel{\rm KLT}{\otimes}$ is the inverse of a matrix whose entries are doubly ordered amplitudes of $\phi^3$. How is the bilinear form modified after including higher derivative corrections in NLSM? Naively, when we include ${\cal O}(p^4)$ corrections to the left-hand side, it is conceivable that there is a version of higher derivative corrections to the cubic bi-adjoint scalar theory that could make the double copy relation work non-trivially. We will see that, for a particular choice of ${\cal O}(p^4)$ operator, there is indeed a theory that would form the double copy relation for $\nlsm^{(4)}$, 
\[
 \nlsm^{(4)} =  \nlsm^{(2)} \stackrel{\rm KLT}{\otimes}\left( {\rm {\textrm{YM}+\phi^3}} \right) \ , 
\]
where YM+$\phi^3$ is a theory of biadjoint scalars with gauge interactions \cite{Chiodaroli:2014xia}.

The paper is organized as follows. In Section \ref{sec:nlsm} we begin with a discussion on the color structure of NLSM at $\mathcal{O}(p^2)$ and $\mathcal{O}(p^4)$. In Section \ref{sec:fk4p} we review the color kinematics duality, including its recent modification, and how it applies to the $\mathcal{O}(p^4)$ single and double trace amplitudes at 4-pt. In Section \ref{sec:fkhm} we then extend to 6-pt, and find a flavor-kinematic solution that matches the $\mathcal{O}(p^4)$ double trace amplitude. In Section \ref{sec:dc} we identify the double copy relation involving the NLSM amplitudes at $\ordr (p^4)$ and YM+$\phi^3$. We end with conclusions and future directions in Section \ref{sec:con}.

\section{The NLSM up to $\ordr (p^4)$}
\label{sec:nlsm}

The NLSM effective Lagrangian can be parameterized as the following:
\bea
\lag_\nlsm = f^2 \Lambda^2 \tilde{\lag} \left(\frac{\partial}{\Lambda}, \frac{\pi}{f} \right),\label{eq:nlsmp}
\eea
where $\pi^a$ are the NGB fields with flavor indices $a$, and $\Lambda$ and $f$ are constants of mass dimension 1, with $f/\Lambda < 1$. The low energy effective Lagrangian is a perturbative expansion of $\partial/\Lambda$, which is predictive when the energy scale of interest is much smaller than $\Lambda$. Because of Lorentz invariance, there are only even powers of $\partial/\Lambda$ in the series expansion when we work in 4 spacetime dimensions.  The leading order, $\ordr (p^2)$ Lagrangian, $\lag_\nlsm^{(2)}$, contains all the terms in Eq. (\ref{eq:nlsmp}) with two derivatives, and the subleading, $\ordr (p^4)$ Lagrangian $\lag_\nlsm^{(4)}$ contains all the terms of four derivatives, and so on. In other words,
\bea
\lag_\nlsm = \lag_\nlsm^{(2)} + \lag_\nlsm^{(4)} +\ordr \left( \frac{1}{\Lambda^4} \right),
\eea
with $\lag_\nlsm^{(2)} = \ordr (1/\Lambda^0)$ and $\lag_\nlsm^{(4)} = \ordr (1/ \Lambda^2)$. At each order in the derivative expansion, the Lagrangian also admits an expansion of $\pi / f$, to all orders in $1/f$.

An $n$-pt tree amplitude has the low energy expansion:
\bea
{\cal M}_n^\nlsm &=& {\cal M}_n^{(2)} + {\cal M}_n^{(4)} + \ordr  \left( \frac{1}{\Lambda^4} \right),\label{eq:dena}
\eea
and at tree level, we have ${\cal M}_n^{(m)} = \ordr (f^{2-n} \Lambda^{2-m} )$. All  vertices in $\lag_\nlsm^{(m)}$ up to $n$-pt will enter ${\cal M}_n^{(m)}$. We  review the Lagrangian and amplitudes of NLSM up to $\ordr (p^4)$ in the following.

\subsection{The Lagrangian}

Let us consider a general NLSM where the NGB fields $\pi^a$ furnish some representation $R$ of a Lie group $H$. Using the bra-ket notation, $|\pi\rangle_a =\pi^a$, the  $\ordr (p^2)$ Lagrangian is \cite{Low:2014nga,Low:2014oga}
\bea
\lag_\nlsm^{(2)} = \frac{f^2}{2} \< d_\mu d^{\mu} \>,\label{eq:glag}
\eea
where
\bea
|d_\mu \> &=&  \frac{1}{f} F_1 (\mt)| \partial_\mu \pi \>,\label{eq:nlddef}\\
F_1 (\mt) &=& \frac{\sin \sqrt{\mt}}{\sqrt{ \mt} } \ ,\\
(\mt)_{ a  b} &=& \frac{1}{f^2} (T^i)_{ a c} (T^i)_{d b}\  \pi^c \pi^{d},
\eea
with $T^i$ being the generators of $H$ in the representation $R$, written in a purely imaginary and anti-symmetric basis: $(T^i)_{ab}=-(T^i)_{ab}^*=-(T^i)_{ba}$. The form of Eq. (\ref{eq:glag}) is fixed by the requirement that the on-shell amplitudes vanish in the single soft limit. This implies a shift symmetry in the Lagrangian \cite{Low:2014nga,Low:2014oga},
\begin{equation}
\label{eq:shiftall}
|\pi \> \to |\pi\> + \sqrt{\mt} \cot \sqrt{\mt}\  |\vep \>\ , 
\end{equation}
where $(|\varepsilon \rangle)_a = \vep^a$ represents an infinitesimal constant ``shift'' in $\pi^a$, as well as a ``closure condition'' that the generators $T^i$ need to satisfy:
\bea
(T^i)_{ab} (T^i)_{cd} + (T^i)_{ac} (T^i)_{db}+(T^i)_{ad} (T^i)_{bc} = 0.\label{eq:clocon}
\eea

Such a condition means that the NLSM can be embedded into a symmetric coset $G/H$. In other words, it can be generated by the spontaneous symmetry breaking of some group $G$, with the coset $G/H$ being symmetric.  The generators of $G$ include the ``unbroken generators'' $\gT^i$ associated with the group $H$, and ``broken generators'' $\gX^a$ associated with the coset $G/H$. Then we can identify $T^i_{ab}  = -if^{iab}$, where $f^{iab}$ is the structure constant of group $G$, so that $[\gX^a, \gX^b] = if^{iab} \gT^i$. Other commutation relations in $G$ include $[\gT^i, \gX^a] = if^{iab} \gX^{b}$ and $[\gT^i, \gT^j] = if^{ijk} \gT^k$, while $ f^{ija} = 0$ as $H$ is a subgroup of $G$, and $f^{abc} = 0$ because we require $G/H$ to be symmetric. Then the Lagrangian can be rewritten as \cite{Coleman:1969sm,Callan:1969sn}
\bea
\lag_\nlsm^{(2)} &=& \frac{f^2}{8} \tr \left( \partial_\mu U^\dagger \partial^\mu U \right),\label{eq:lagst} \qquad
U = \exp \left(2i \pi^a \gX^a /f \right).
\eea
The interactions given by Eq. (\ref{eq:lagst}) are even powers of $\pi^a$ contracted with a single trace of generators $\gX^a$.

The Lagrangian at the subleading order of $\ordr (p^4)$ in general contains four independent Parity-even operators:
\bea
\lag_\nlsm^{(4)} = \frac{f^2}{\Lambda^2}  \sum_{i=1}^4 C_i O_i ,\label{eq:nlmgutp4}
\eea
where
\bea
O_1  =  [ \tr ( d_\mu d^\mu ) ]^2, \ O_2  =  [\tr ( d_\mu d_\nu ) ]^2, \ O_3  =  \tr ( [ d_\mu, d_\nu]^2 ),\ O_4  =  \tr ( \{ d_\mu, d_\nu \}^2 ),\label{eq:op4o}
\eea
and $d_\mu = d_\mu^a \gX^a$. In $4$ spacetime dimensions, there can also be a Wess-Zumino-Witten term \cite{Wess:1971yu,Witten:1983tw}, which we will not consider for now.

\subsection{Flavor ordering of the amplitudes}
\label{sec:rb}

The on-shell method to construct the NLSM interactions for a general symmetric coset is soft bootstrap \cite{Cheung:2015ota,Elvang:2018dco,Low:2019ynd}, where we consider flavor-ordered partial amplitudes. For the NLSM at $\ordr (p^2)$, which we will denote as $\nlsm^{(2)}$, the partial amplitudes are similar to the color-ordered amplitudes of the YM theory \cite{Dixon:1996wi}, where the interactions involve the structure constant $f^{ijk}$, which for NLSM can be identified with $(T^i)_{ab} $ as in Eq. (\ref{eq:glag}). From the perspective of the unbroken group $H$, $(T^i)_{ab}$ is a group generator in some general representation; however, from the perspective of broken group $G$ and coset $G/H$, $(T^i)_{ab} = -if^{iab}$ is the structure constant of $G$, i.e. the generator of $G$ in the adjoint representation. Similarly, the gauge bosons in YM theories furnish the adjoint representation as well.

Therefore, the color-decomposition of YM theories can be directly applied to general $\nlsm^{(2)}$. The flavor structure of the full amplitude can be expanded in the trace basis as
\bea
{\cal M}_n^{(2),a_1 \cdots a_n} (p_1, \cdots, p_n) = \sum_{\alpha \in S_{n-1}} \tr \left( \gX^{a_1} \gX^{a_{\alpha (1)}} \cdots \gX^{a_{\alpha (n-1)}} \right) M_n^{(2)} (1,\alpha),\label{eq:fdtb}
\eea
where $\alpha$ is a permutation of $\{2, 3, \cdots, n\}$ and $M_n (1,\alpha)$ is the single-trace flavor-ordered amplitude. The right-hand side of Eq. (\ref{eq:fdtb}) is a sum of $(n-1)!$ terms.

The lesson we learn from YM theories is that the flavor expansion in Eq. (\ref{eq:fdtb}) is over-complete, and can be further reduced to the  Del Duca-Dixon-Maltoni (DDM) basis \cite{DelDuca:1999rs} as a sum of $(n-2)!$ terms:
\bea
&&{\cal M}_n^{(2),a_1 \cdots a_n} (p_1, \cdots, p_n)\non\\
 &=& \sum_{\alpha \in S_{n-2}} (-1)^{n/2-1} f^{a_1 a_{\alpha (1)} i_1 }\left( \prod_{j=1}^{n/2-2} f^{i_{j} a_{ \alpha (2j)} b_j} f^{b_j a_{\alpha (2j+1)} i_{j+1}} \right)\non\\
&&\times  f^{i_{n/2-1} a_{\alpha (n-2)} a_n} M_n^{(2)} (1,\alpha, n)\non\\
&=& \sum_{\alpha \in S_{n-2}}  T^{i_1}_{a_1 a_{\alpha (1)}  }\left( \prod_{j=1}^{n/2-2} T^{i_{j}}_{ a_{ \alpha (2j)} b_j} T^{i_{j+1}}_{b_j a_{\alpha (2j+1)} } \right)  T^{i_{n/2-1}}_{ a_{\alpha (n-2)} a_n} M_n^{(2)} (1,\alpha, n),\label{eq:ddm}
\eea
where $\alpha $ is a permutation of $\{2,3,\cdots, n-1\}$. This implies the  flavor-ordered amplitudes need to satisfy $(n-1)!-(n-2)!$  Kleiss-Kuijf (KK)  relations \cite{Kleiss:1988ne}. The true number of independent flavor-ordered amplitudes is further reduced to $(n-3)!$  by the BCJ relations \cite{Bern:2008qj}.

\begin{figure}[htbp]
\centering
\includegraphics[width=0.5\textwidth]{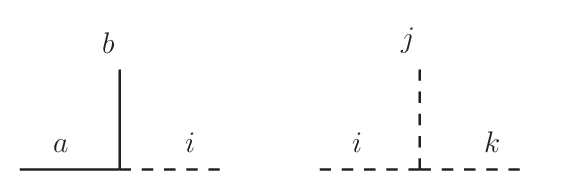}
\caption{\label{fig:tf} The graphic presentation of $(T^{i})_{ab}$ and respectively $f^{ijk}$ as vertices.}
\end{figure}

\begin{figure}[htbp]
\centering
\includegraphics[width=0.7\textwidth]{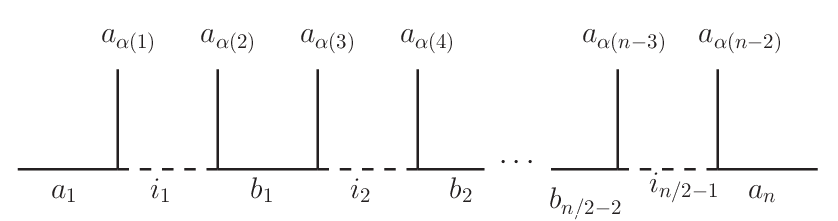}
\caption{\label{fig:ddm} The flavor factors in the DDM basis. The internal lines represent the indices that are contracted and summed over.}
\end{figure}

We can express the flavor factors in Eq. (\ref{eq:ddm}) diagrammatically, where the broken indices $a$ and the unbroken indices $i$ are represented by solid and dashed lines, respectively.  The generator $(T^i)_{ab}$ and the structure constant $f^{ijk}$ are then vertices given by Fig. (\ref{fig:tf}). Under this notation, the flavor factor of each term in Eq. (\ref{eq:ddm}) is given by a half-ladder graph shown in Fig. \ref{fig:ddm}.

For amplitudes in the higher orders of the derivative expansion, flavor structures which are products of multiple traces can appear. The most general multi-trace flavor decomposition is the following:
\begin{align}
\ \ \ \ \M^{a_1 \cdots a_{n} }_n (p_1, \cdots, p_{n})& \equiv&
\sum_{t=1}^{\lfloor n/2 \rfloor } \sum_{ l}  \sum_{\sigma \in S_{n} / S_{n;l} } \left[ \prod_{i=1}^{t} \tr \left( \gX^{a_{\sigma  (l_{i-1} +1)}} \cdots \gX^{ a_{\sigma (l_i)}}  \right) \right]  M_{\sigma ; l} (p_1, \cdots , p_n)\, ,\label{eq:mttfo}
\end{align}
where $l = \{ l_0, \cdots ,l_t \}$ labels possible partition of ordered indices $\{1, 2, \cdots , n\}$ into $t$ subsets, with the requirement of $l_0 = 0$, $l_t = n$ and $l_{i+1} - l_i \le l_{i+2} - l_{i+1}$, $i=0,1, \cdots, t-2$; $S_{n;l}$ are the permutations of $\{1, 2, \cdots , n\}$ that leave the flavor factor invariant. We can denote $M_{\sigma ; l} (p_1, \cdots , p_n)$ as
\bea
M_n ( \sigma (1), \cdots, \sigma (l_1) | \sigma(l_1 + 1), \cdots , \sigma (l_2 ) | \cdots | \sigma (l_{t-1} + 1), \cdots \sigma (n) ).
\eea
The amplitude
\bea
M_n (1,2, \cdots l_1|l_1 + 1, \cdots , l_2 | \cdots | l_{t-1} + 1, \cdots, n)
\eea
is invariant when we  do the cyclic permutations separately for the sets of indices $\{1,2, \cdots, l_1\}$, $\{ l_1+1, \cdots, l_2 \}$ and so on. Furthermore, if $l_{i+1} - l_i = l_{i+2} - l_{i+1}$, exchanging the sets $\{ l_i+1, \cdots ,l_{i+1} \}$ and $\{ l_{i+1} + 1, \cdots , l_{i+2} \}$ will also leave the amplitude invariant, as the flavor factors associated with the amplitudes remain the same. In general, the $\ordr (p^4)$ tree amplitude $\M_n^{(4)}$ can only have single and double trace flavor factors, which is easy to see from the observation that at ${\cal O}(p^4)$ NLSM Lagrangian contains only single- and double-trace operators.

\section{Flavor-kinematics duality at $4$-pt}

\label{sec:fk4p}

The color-kinematics duality of scattering amplitudes was first discovered for YM theories, the $n$-pt tree amplitudes of which can be written in the following form \cite{Bern:2008qj,Bern:2019prr}:
\bea
{\cal M}_n^{\text{YM}} = \sum_{g \in \{g_n\!\} } \frac{\mathsf{c}_g\ n_g}{d_g},\label{eq:ckym}
\eea
where the sum is over all distinct $n-$pt cubic graphs $\{ g_n \}$, while $\mathsf{c}_g$, $n_g$ and $d_g$ are the color numerators, kinematic numerators and denominators of each cubic graph. The denominators $d_g$ are given by the propagators associated with the cubic graphs, $n_g$ only contains kinematic information (Mandelstam invariants and polarization vectors), while  the color structures are isolated in $\mathsf{c}_g$. The gauge fields are in the adjoint representation, and $\mathsf{c}_g$ are constructed using structure constants, thus they satisfy anti-symmetry and the Jacobi identity. The duality for color and kinematics manifests in the fact that it is possible to find a representation for $n_g$ so that they satisfy anti-symmetry and the Jacobi identity as well.

It is known that such a duality also exists for the tree amplitudes of $\nlsm^{(2)}$ \cite{Chen:2013fya,Du:2016tbc,Carrasco:2016ldy}. However, as an EFT by construction, NLSM admits a derivative expansion, as shown in Eq. (\ref{eq:dena}). \textit{A priori} it is not clear whether the higher order contributions in the derivative expansion have the same property as well. For the next-to-leading order, i.e. $\ordr (p^4)$, previous works of directly applying the ${\cal O}(p^2)$ BCJ relations fail to hold \cite{Elvang:2018dco,Carrillo-Gonzalez:2019aao}. It turns out that, for the flavor-kinematics duality to work at $\ordr (p^4)$, we need to generalize our definitions for the color/flavor numerators. 

\subsection{Building $4$-pt numerators}

We start with the lowest multiplicity, which is $n=4$ for the NLSM amplitudes. Recently new ways to construct 4-pt numerators have been proposed \cite{Carrasco:2019yyn,Low:2019wuv}, and we will discuss them systematically in the following.

\begin{figure}[htbp]
\centering
\includegraphics[width=0.3\textwidth]{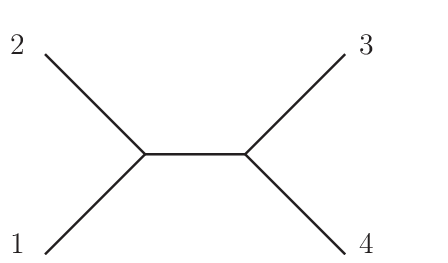}
\caption{The $s$-channel 4-pt cubic graph.\label{fig:4ptcub}}
\end{figure}
At 4-pt, we can define a function with three indices $\sj(1,2,3)$ and associate it with the 4-pt cubic graph in Fig. \ref{fig:4ptcub}. If we impose anti-symmetry and the Jacobi identity, then it can be used as the numerator for 4-pt amplitudes. Specifically, we want
\bea
\sj(1,2,3) = -\sj(2,1,3),\quad  \sj (1,2,3) + \sj(2,3,1) + \sj(3,1,2) = 0. \label{eq:definingre}
\eea
It is also convenient to define
\bea
\sj_s  \equiv \sj (1,2,3),\ \sj_t \equiv \sj (2,3,1),\ \sj_u \equiv \sj (3,1,2),
\eea
with
\bea
s \equiv s_{12}, \ t \equiv s_{23}, \ u\equiv s_{13},
\eea
satisfying $s + t + u = 0$, where $s_{ij} \equiv (p_i + p_j)^2$.

Let us first consider color/flavor numerators that do not contain any kinematic information. For YM, the color numerator is given by
\bea
\mathsf{c} (1,2,3) = f^{ia_1 a_2} f^{ia_3 a_4},
\eea
where $f^{iab}$ are structure constants of some Lie group $H$. More generally, for generators of $H$ in some representation $R$, this can be generalized to
\bea
\sff_R (1,2,3) = (T^i)_{a_1 a_2} (T^i)_{a_3 a_4} \ , \label{eq:frdef}
\eea
assuming the closure condition given by Eq. (\ref{eq:clocon}). This is the flavor numerator for an $\nlsm^{(2)}$ amplitude for a general group $H$. The color numerator $\mathsf{c} = \sff_A$ is just a special case where the representation is the adjoint $A$. 

Another valid flavor numerator is given by
\bea
\sff_\delta (1,2,3) = \delta^{a_1 a_3} \delta^{a_2 a_4} - \delta^{a_1 a_4} \delta^{a_2 a_3},
\eea
where the indices can be in any representation for any group: $\delta^{ab}$ is always an invariant tensor. It is easy to check that $\sff_\delta$  satisfies anti-symmetry and the Jacobi identity. One can also identify that
\bea
\sff_{\delta} (1,2,3) \propto \sff_{R} (1,2,3) \label{eq:fpfso}
\eea
when $R$ is the fundamental representation of $\SO (N)$. The above is the consequence of the completeness relations of the generators in the fundamental representation of $\SO (N)$:
\bea
(T^i)_{ab } (T^i)_{cd} = \frac{1}{2}  (\delta^{ad} \delta^{bc} - \delta^{ac} \delta^{bd})\ ,\label{eq:comrel}
\eea
and it is known that NLSM of fundamental $\SO (N)$ can be embedded to the symmetric coset $\SO (N+1) / \SO (N)$. In other words, for the $\SO (N)$ fundamental representation, $\sff_{\delta}$ is not a new building block but is identical to $\sff_{R}$. For other group representations, it is indeed new.

Next, let us consider numerators containing kinematic invariants. For simplicity we will restrict ourselves to numerators that are local. At the lowest mass dimension, we have the following numerator that only contains momenta:
\bea
n^{\ss} (1,2,3) = t-u,
\eea
which is the kinematic numerator for single-flavor YM scalar theory.

We can use the simple building blocks discussed in the above to construct more complicated numerator $\sj$'s. One way is to just multiply existing numerators with permutation invariant objects. There are  two such objects that encode  the internal symmetry:
\bea
\begin{split}
\sd^{ abcd}_4 &=  \sum_{\sigma \in S_3} \tr \left( T^{a_{\sigma (1)}} T^{a_{\sigma (2)}} T^{a_{\sigma (3)}} T^{a_4} \right)\ ,\\
 \sd^{ abcd}_2 &= \frac{1}{2}  \sum_{\sigma \in S_3} \delta^{a_{\sigma (1)} a_{\sigma (2)}} \delta^{a_{\sigma (3)}a_4} .
 \end{split}
\eea
For non-adjoint representations, $\sd_4$ can be generalized to any rank-4 totally symmetric tensor $\sd_4$, which may or may not  exist. There are also two permutation invariant building blocks that only contain kinematic invariants:
\bea
\sX \equiv stu,\ \qquad \qquad \sY \equiv  s^2 + t^2 + u^2. \label{eq:XYdef}
\eea

The other way to generate new numerators is to take two existing numerators $\sj$ and $\sj'$, and define $\sJ (\sj,\sj') = \sj_t \sj_t' - \sj_u \sj_u'$. Then $\sj'' (1,2,3) = \sJ (\sj,\sj')$ is a perfectly valid new numerator.

Now let us build more numerators only containing momenta. We have
\bea
\sJ (n^\ss, n^\ss) \propto  n^\nl (1,2,3) = \frac{1}{3} s(t-u), \label{eq:defnnl}
\eea
which is the kinematic numerator for $\nlsm^{(2)}$ \cite{Du:2016tbc,Carrasco:2016ldy}. We also have
\bea
\label{comp2}
\sJ ( n^\ss , n^\nl) = \frac{1}{6} \sY n^\ss_s,\ \qquad \qquad \sJ(n^\nl, n^\nl) =- \frac{1}{6} (\sX n^\ss_s + \sY n^\nl_s).
\eea
This means that all numerators that only contain momenta can be written as a linear combination of $n^\ss$ and $n^\nl$, each dressed with powers of permutation invariant objects $\sX$ and $\sY$ \cite{Carrasco:2019qwr}.

\subsection{$4$-pt soft blocks at $\ordr (p^4)$ for NLSM}

\label{sec:sb4pt}

The full 4-pt amplitude for $\nlsm^{(2)}$ is \cite{Bern:2008qj,Bern:2019prr} 
\bea
{\cal M}^{(2)}_4 = \frac{1}{f^2} \left(\frac{\sff_{R,s}\ n^\nl_s}{s} +\frac{\sff_{R,t}\ n^\nl_t}{t} + \frac{\sff_{R,u}\ n^\nl_u}{u} \right),\label{eq:nlfa2}
\eea
where we have suppressed the flavor indices. Here $\sff_{R,s/t/u}$ is the flavor factor defined in Eq.~(\ref{eq:frdef}) for the $s/t/u$ channel, while  $n^\nl$ is defined in Eq.~(\ref{eq:defnnl}).
 We can rearrange the amplitude to the DDM basis \cite{DelDuca:1999rs} using $\sff_{R,t}  = -\sff_{R,s}- \sff_{R,u}$:
\bea
{\cal M}^{(2)}_4 = \frac{1}{f^2} \left[ \sff_{R,s} \left(\frac{ n^\nl_s}{s} - \frac{ n^\nl_t}{t} \right) + \left(-  \sff_{R,u} \right) \left(\frac{ n^\nl_t}{t} - \frac{ n^\nl_u}{u} \right) \right], \label{eq:m24ddm}
\eea
where the two terms correspond to the flavor-ordered partial amplitudes:
\bea
\begin{split}
M_4^{ (2)} (1,2,3,4) &=\frac{1}{f^2}\left( \frac{ n^\nl_s}{s} - \frac{ n^\nl_t}{t} \right) = -\frac{1}{f^2} u \ , \\
M_4^{ (2)} (1,3,2,4) &=\frac{1}{f^2}\left(  \frac{ n^\nl_t}{t} - \frac{ n^\nl_u}{u} \right)= -\frac{1}{f^2} s\ . \label{eq:m41234}
\end{split}
\eea

Now let us consider the $\ordr (p^4)$ contributions in the NLSM, which in general are given by the operators $O_{1,\cdots,4}$ in Eq. (\ref{eq:op4o}). Their contributions to the 4-pt flavor-ordered amplitudes in the trace basis can be characterized by the following 4 soft blocks \cite{Low:2019ynd}:
\begin{align}
& &\text{Single-trace:}\ \ {\cal S}_1^{(4)} (1,2,3,4) &=  \frac{1}{\Lambda^2 f^2} \  u^2 \ , \   & {\cal S}_2^{(4)} (1,2,3,4)&=   \frac{1}{\Lambda^2 f^2} \  st\ ,\label{eq:sa4s}\\
&&\text{Double-trace:}\ \ {\cal S}_1^{(4)} (1,2|3,4) &=   \frac{1}{\Lambda^2 f^2}  \ s^2 \ , \   & {\cal S}_2^{(4)} (1,2|3,4)&=    \frac{1}{\Lambda^2 f^2} \ tu\ .
\end{align}
The full 4-pt amplitude can be written as
\bea
{\cal M}_4 &=& {\cal M}^{(2)}_4 +  \sum_{\sigma \in S_3} \tr \left( T^{a_{\sigma (1)}} T^{a_{\sigma (2)}} T^{a_{\sigma (3)}} T^{a_4} \right) \left[ c_1\ {\cal S}_1^{(4)} (\sigma,4) + c_2\ {\cal S}_2^{(4)} (\sigma,4)  \right] \non\\
&&+ \frac{1}{2}  \sum_{\sigma \in S_3} \delta^{a_{\sigma (1)} a_{\sigma (2)}} \delta^{a_{\sigma (3)}a_4} \left[ d_1\ {\cal S}_1^{(4)} (\sigma (1), \sigma (2) |\sigma (3) ,4) + d_2\ {\cal S}_2^{(4)} (\sigma (1), \sigma (2) |\sigma (3) ,4)\right]\non\\
&& + \ordr \left( \frac{1}{\Lambda^4} \right),
\eea
where the dimensionless constants $c_i$ and $d_i$ are related to the Wilson coefficients in Eq. (\ref{eq:nlmgutp4}) by
\bea
 c_1 =C_3+ 3 C_4 ,\qquad c_2 = 2(C_3- C_4),\qquad d_1 &=& 2C_1 + C_2,\qquad d_2 = 2C_2.\label{eq:lacr}
\eea

Now we want to construct a 4-pt amplitude at $\ordr (p^4)$ that is local and exhibits flavor-kinematics duality, using numerators that satisfy the anti-symmetry and Jacobi identity in Eq.~(\ref{eq:definingre}). One natural possibility is to replace the ${\cal O}(p^2)$ kinematic numerator $n^\nl$ in Eq.~(\ref{eq:nlfa2}) with  ${\cal O}(p^4)$ kinematic invariants while leaving the flavor factors $\sff_{R}$ intact. In this case the   coefficients of $\sff_{R}$ in the DDM basis  simply correspond to flavor-ordered partial amplitudes, cf.~Eq.~(\ref{eq:m24ddm}), and satisfy KK and BCJ relations. However, it was shown in Refs.~\cite{Elvang:2018dco,Carrillo-Gonzalez:2019aao} that such local kinematic numerators do not exist at ${\cal O}(p^4)$.

An alternative possibility is to leave the ${\cal O}(p^2)$ kinematic numerator $n^\nl$ unchanged and modify the flavor factor $\sff\to \hat{\sff}$,
\bea
{\cal M}^{(4)}_4 = \frac{1}{f^2} \left(\frac{\hat{\sff}_s\ n^\nl_s}{s} +\frac{\hat{\sff}_t\ n^\nl_t}{t} + \frac{\hat{\sff}_u\ n^\nl_u}{u} \right)\ ,
\eea
where  $\hat{\sff}$ is now ${\cal O}(p^2)$ in order for the full amplitudes to be ${\cal O}(p^4)$. Assuming the flavor-kinematics duality, $\hat{\sff}_u= -\hat{\sff}_t-\hat{\sff}_s$, the full amplitude can be written as
\bea
{\cal M}^{(4)}_4=\frac{1}{f^2} \left( \hat{\sff}_s\, M_4^{ (2)} (1,2,3,4) - \hat{\sff}_u \, M_4^{ (2)} (1,3,2,4) \right)\ , \label{eq:nlfa4}
\eea
where we have plugged in Eq.~(\ref{eq:m41234}). We see the ansatz in Eq.~(\ref{eq:nlfa4}) amounts to expanding the ${\cal O}(p^4)$ full amplitudes in terms of ${\cal O}(p^2)$ partial amplitudes. We will present four different possibilities for $\hat{\sff}$.

The first possibility is
\bea
\hat{\sff}_1 (1,2,3) = \frac{1}{\Lambda^2} \sJ (\sff_{R},n^\ss) = \frac{1}{\Lambda^2}\left[ \sff_{R,t} (u-s) - \sff_{R,u} (s-t) \right].\label{eq:fh1}
\eea
This gives us a local full amplitude, and we can rewrite it in the DDM basis:
\bea
{\cal M}^{(4)}_{\hat{\sff}_1,4} = -\frac{1}{\Lambda^2 f^2 } \left( \hat{\sff}_{1,s} u - \hat{\sff}_{1,u} s \right)= -\frac{1}{\Lambda^2 f^2} \left[ \sff_{R,s} (-u^2 - 2st) +(-\sff_{R,u}) (-s^2 - 2tu) \right].
\eea
Therefore, we obtain a partial amplitude in the single-trace basis:
\bea
M^{(4)}_{\hat{\sff}_1,4} (1,2,3,4) = \frac{1}{\Lambda^2 f^2} (u^2 + 2st) =  {\cal S}_1^{(4)} (1,2,3,4) + 2 {\cal S}_2^{(4)} (1,2,3,4),
\eea
which is the unique single-trace soft block at 4-pt that satisfies KK relations \cite{Kleiss:1988ne}.

The second modified flavor numerator is
\bea
\hat{\sff}_2 (1,2,3)  =\frac{1}{\Lambda^2} \sd^{ a_1 a_2 a_3 a_4}_4 n^\ss_s = \frac{1}{\Lambda^2}  \sd^{ a_1 a_2 a_3 a_4}_4 (t-u).\label{eq:fh2}
\eea
Again, the corresponding full amplitude is local, while this time we write it in the trace basis:
\bea
{\cal M}^{(4)}_{\hat{\sff}_2,4}  &=& \frac{6}{\Lambda^2 f^2 } \sd^{ a_1 a_2 a_3 a_4} \sY = \frac{1}{ \Lambda^2 f^2 } \sum_{\sigma \in S_3} \tr \left( T^{a_{\sigma (1)}} T^{a_{\sigma (2)}} T^{a_{\sigma (3)}} T^{a_4} \right) \sY.
\eea
Then the partial amplitude is
\bea
M^{(4)}_{\hat{\sff}_2,4} = \frac{1}{ \Lambda^2 f^2 } \sY = \frac{2}{ \Lambda^2 f^2 } (u^2 - st)= 2\left[ {\cal S}_1^{(4)} (1,2,3,4)  - {\cal S}_2^{(4)} (1,2,3,4) \right].
\eea
This is the unique single-trace soft block at 4-pt that is permutation invariant.

To obtain flavor-ordered partial amplitudes corresponding to the two double trace soft blocks we just need to replace $\sff_{R}$ in Eq. (\ref{eq:fh1}) with $\sff_{\delta}$, and $\sd_4$ in Eq. (\ref{eq:fh2}) with $\sd_2$. Recall that the full double-trace amplitude is given by
\bea
{\cal M}_4  = \delta^{a_1 a_2} \delta^{a_3 a_4} M_4 (1,2|3,4) + \delta^{a_1 a_3} \delta^{a_2 a_4} M_4 (1,3|2,4) + \delta^{a_1 a_4} \delta^{a_2 a_3} M_4 (1,4|2,3).
\eea
Let us first replace $\sff_{R}$ in Eq. (\ref{eq:fh1}) with $\sff_{\delta}$:
\bea
\hat{\sff}_3 (1,2,3) = \frac{1}{\Lambda^2} \sJ (\sff_{\delta},n^\ss) = \frac{1}{\Lambda^2}\left[ \sff_{\delta,t} (u-s) - \sff_{\delta,u} (s-t) \right].
\eea
The corresponding full amplitude is
\bea
{\cal M}^{(4)}_{\hat{\sff}_3,4} &=& -\frac{1}{\Lambda^2 f^2} \left[ \sff_{\delta,s} (-u^2 - 2st) +(-\sff_{\delta,u}) (-s^2 - 2tu) \right] \non\\
&=&\frac{1}{\Lambda^2 f^2} \left[ \delta^{a_1 a_2} \delta^{a_3 a_4} (s^2 + 2tu) + \delta^{a_1 a_3} \delta^{a_2 a_4} (u^2 + 2st)   + \delta^{a_1 a_4} \delta^{a_2 a_3}(t^2 + 2su)  \right],
\eea
which gives the partial amplitude
\bea
 M^{(4)}_{\hat{\sff}_3,4} (1,2|3,4) = \frac{1}{\Lambda^2 f^2} (s^2 + 2tu) =  {\cal S}_1^{(4)} (1,2|3,4) + 2 {\cal S}_2^{(4)} (1,2|3,4).\label{eq:fk4pf3}
\eea
On the other hand, replacing $\sd_4$ in Eq. (\ref{eq:fh2}) with $\sd_2$ leads to
\bea
\hat{\sff}_4 (1,2,3)  =\frac{1}{\Lambda^2} \sd_2^{ a_1 a_2 a_3 a_4} n^\ss_s = \frac{1}{\Lambda^2} \sd_2^{ a_1 a_2 a_3 a_4} (t-u),
\eea
so that the full amplitude becomes
\bea
{\cal M}^{(4)}_{\hat{\sff}_4,4}  = \frac{1}{2 \Lambda^2 f^2 } \sum_{\sigma \in S_3} \delta^{a_{\sigma (1)} a_{\sigma (2)}} \delta^{a_{\sigma (3)}a_4} \sY = \frac{1}{ \Lambda^2 f^2 } \left[ \delta^{a_1 a_2} \delta^{a_3 a_4}  + \delta^{a_1 a_3} \delta^{a_2 a_4}  + \delta^{a_1 a_4} \delta^{a_2 a_3} \right]\sY \ ,.
\eea
where $\sY$  is defined in Eq.~(\ref{eq:XYdef}).
Then the partial amplitude is
\bea
M^{(4)}_{\hat{\sff}_4,4} (1,2|3,4) = \frac{1}{ \Lambda^2 f^2 } \sY = \frac{2}{ \Lambda^2 f^2 } (s^2 -tu) = 2 \left[{\cal S}_1^{(4)} (1,2|3,4)  - {\cal S}_2^{(4)} (1,2|3,4) \right].\label{eq:fk4pf4}
\eea

In the end, the four different modified flavor factors give rise to flavor-ordered partial amplitudes corresponding to the four soft blocks.

\section{Flavor-kinematics duality at higher multiplicity}

\label{sec:fkhm}

Inspired by the 4-pt results discussed in the last section, we assume the following ansatz  for the $n$-pt full amplitude of NLSM at $\ordr (p^4)$, 
\bea
{\cal M}_{\hat{\sff}_i,n}^{(4)} = \frac{1}{f^{n-2}} \sum_{g \in \{g_n\} } \frac{\hat{\sff}_{i,g}\ n_{g}^\nl}{d_{g}},
\eea
where $d_{g} = \ordr ( p^{2n-6})$, $n_{g}^\nl = \ordr ( p^{2n-4})$, and $\hat{\sff}_{i,g} = \ordr ( p^2)$. Again using the Jacobi relations among $\hat{\sff}_{i,g}$ the full amplitude in the DDM basis is an expansion in the $\nlsm^{(2)}$ partial amplitudes $M_n^{ (2)}$:
\bea
{\cal M}_{\hat{\sff}_i,n}^{(4)} = \sum_{\sigma \in S_{n-2}}\hat{\sff}_{i, \hl(1,\sigma,n)}\ M_n^{ (2)} (1, \sigma, n),\label{eq:npd}
\eea
where $\sigma$ is a permutation of $\{ 2, 3, \cdots, n-1\}$, $\hl(1,\sigma,n)$ is the corresponding half-ladder graph with $1$ and $n$ at two ends, as shown in Fig. \ref{fig:ddm}. On the other hand, we can also expand ${\cal M}_{\hat{\sff}_i,n}^{(4)}$ in the trace basis:
\bea
{\cal M}_{\hat{\sff}_i,n}^{(4)} = \sum_{\sigma} f_{i,\sigma}\ M_{\hat{\sff}_i,n}^{(4)} (\sigma),\label{eq:npt}
\eea
where $\sigma$ corresponds to all the distinct trace structures, $f_{i,\sigma}$ is the flavor factor which is either a single trace or a product of two traces, and $M_{\hat{\sff}_i,n}^{(4)}$ is the flavor-ordered amplitude of the $\nlsm^{(4)}$ in the trace decomposition. That both Eq. (\ref{eq:npd}) and Eq. (\ref{eq:npt}) are true means that
\bea
M_{\hat{\sff}_i,n}^{(4)} (\mathbb{I}_n ) = \sum_{\sigma \in S_{n-2}} c_\sigma^{(i)}\ M_n^{ (2)} (1, \sigma, n),\label{eq:exp}
\eea
where $c_\sigma^{(i)} = \ordr ( p^2)$ is a function of the momentum invariants. In other words, the $\ordr (p^4)$ flavor-ordered amplitudes can be expanded in terms of the $\ordr (p^2)$ flavor-ordered amplitudes; the expansion coefficients come from $\hat{\sff}_{i}$ and are at $\ordr (p^2)$.

Therefore, to find $\hat{\sff}_{i}$ for higher multiplicity amplitudes, all we need to do is solve the coefficients $c_\sigma^{(i)}$ in Eq.~(\ref{eq:exp}), plug in the solution to Eq.~(\ref{eq:npt}), and use Eq.~(\ref{eq:npd}) to find $\hat{\sff}_{i, \hl(1,\sigma,n)}$. Once this is done, all the other $\hat{\sff}_{i,g}$  can be uniquely determined using the Jacobi relations. In the most general case, without imposing any constraints on $c_\sigma^{(i)}$,  the number of solutions are infinite. Instead, we will assume that $c_\sigma^{(i)}$, and consequently $\hat{\sff}_{i,g}$, are local, and proceed to look for solutions.

Let us illustrate this in the simplest case of 4-pt amplitudes. Take the single trace amplitude
\bea
M^{(4)}_{\hat{\sff}_2,4} (1,2,3,4) = \frac{2}{\Lambda^2 f^2} (u^2 - st) 
\eea
as an example, and expand it as in Eq. (\ref{eq:exp}). The two $\nlsm^{(2)}$ amplitudes we need are $M_4^{ (2)} (1,2,3,4)$ and $M_4^{(2)} (1,3,2,4)$, which are given in Eq. (\ref{eq:m41234}). The solutions for $c_\sigma^{(2)}$ in
\bea
M_{\hat{\sff}_2,4}^{(4)} (1,2,3,4 ) = \sum_{\sigma \in S_{2}} c_\sigma^{(2)} M_4^{ (2)} (1, \sigma, 4)\label{eq:exp4p}
\eea
is
\bea
c_{2,3}^{(2)} = -\frac{1}{\Lambda^2} \left[2u + (1+\alpha) s \right], \qquad c_{3,2}^{(2)} =- \frac{1}{\Lambda^2} \left[2s +(1-\alpha) u \right],\label{eq:coes4p}
\eea
where $\alpha$ is an arbitrary constant. The degree of freedom in the solution, which is characterized by $\alpha$, is a consequence of the single BCJ relation of $M^{(2)}_4$:
\bea
s M_4^{ (2)} (1,2,3,4) - u M_4^{(2)} (1,3,2,4) = 0.
\eea
We can then work out the modified flavor factor $\hat{\sff}_2$ by plugging Eq. (\ref{eq:exp4p}) into the full amplitude:
\bea
\M_{\hat{\sff}_2,4}^{(4)} &=& \sum_{\sigma \in S_3} \tr \left(1 \sigma  \right)\ M_{\hat{\sff}_1,4}^{(4)} (1, \sigma)\non\\
&=&   \hat{\sff}_{2,s} (\alpha)\, M_4^{ (2)} (1,2,3,4) - \hat{\sff}_{2,u}(\alpha)\, M_4^{ (2)} (1,3,2,4) ,\label{eq:4ptfc2}
\eea
with
\bea
\hat{\sff}_{2,s} (\alpha) &=& \frac{2}{\Lambda^2} \left\{ \tr (1234) \left[(1+\alpha)t-(1-\alpha)u \right]\right. \non\\
&&\left.+ \left[ \tr (1324) + \tr (1342)  \right] \left[(1-\alpha)t - (1+\alpha) u \right]\right\},\label{eq:f2ffgs}\\
\hat{\sff}_{2,u} (\alpha) &=& \frac{2}{\Lambda^2} \left\{ \tr (1234) \left[(1+\alpha)s-(1-\alpha)t \right]\right. \non\\
&&\left.+ \left[ \tr (1324) + \tr (1342)  \right] \left[(1-\alpha)s - (1+\alpha) t \right]\right\},
\label{eq:f2ffgu}
\eea
where we have used the shorthand notation for the traces:
\bea
\tr (123\cdots) \equiv \tr \left( \gX^{a_1} \gX^{a_2 } \gX^{a_3 } \cdots \right).
\eea
To arrive at Eq. (\ref{eq:4ptfc2}) we have used the cyclic and reverse ordering invariance of $\tr (1234)$ and $M_4^{(2)} (1234)$, as well as the 4-pt KK relation for $M_4^{(2)}$:
\bea
M_4^{(2)} (1342) =-M_4^{(2)} (1234) - M_4^{(2)} (1324) .
\eea
Eqs. (\ref{eq:f2ffgs}) and (\ref{eq:f2ffgu}) give the most general modified flavor factors that work for  $\M_{\hat{\sff}_2,4}^{(4)}$. If we also want them to be relabeling symmetric, in this case exchanging $2 \leftrightarrow 3$ resulting in $\hat{\sff}_{2,s} \leftrightarrow -\hat{\sff}_{2,u}$, the constant $\alpha$ must be set to $0$: we have
\bea
\hat{\sff}_{2,s} (0) = \frac{1}{\Lambda^2}  \sd^{ a_1 a_2 a_3 a_4}_4 (t-u),
\eea
which is exactly what we know from Eq. (\ref{eq:fh2}).

At 6-pt there are no local solutions of $c_\sigma$ when all four ${\cal O}(p^4)$ operators are turned on with their respective (arbitrary) Wilson coefficients. We do find a solution, however, for  the 6-pt, $\ordr (p^4)$ amplitude that is soft-bootstrapped from the soft block ${\cal S}_2^{(4)} (1,2|3,4)$, which is written as
\bea
M_{d_2, 6}^{ (4)} (1,2|3,4,5,6)&=& \frac{1}{\Lambda^2f^4} \left[ s_{46} \left(\frac{s_{13} s_{23}}{P^2_{123}} + \frac{s_{15} s_{25}}{P^2_{125}}  \right)  + s_{35} \left(\frac{s_{14} s_{24}}{P^2_{124}} + \frac{s_{16} s_{26}}{P^2_{126}}  \right) \right.\non\\
 &&\left.\phantom{s_{46} \left(\frac{s_{13} s_{23}}{P^2_{123}} \right)}- (s_{15} + s_{13}) (s_{25} + s_{23}) + s_{12} s_{35}  \right],\label{eq:d26p}
\eea
where $P^2_{ijk \cdots} \equiv (p_i + p_j + p_k + \cdots)^2$. From Eq. (\ref{eq:lacr}) we see that this contribution corresponds to the following values for the Wilson coefficients in the Lagrangian:
\bea
C_1 = \frac{1}{4}, \quad C_2 = -\frac{1}{2}, \quad C_3 = C_4 = 0.
\eea
We will denote such a theory as $\nlsm^{d_2}$. At 4-pt, we learned from Eqs. (\ref{eq:fk4pf3}) and (\ref{eq:fk4pf4}) that the associated modified flavor factor which gives this contribution is $(2\hat{\sff}_3 - \hat{\sff}_4)/6$. Eq. (\ref{eq:d26p}) can be expanded in the form of Eq. (\ref{eq:exp}):
\bea
M_{d_2, 6}^{ (4)} (1,2|3,4,5,6) = \sum_{\sigma \in S_{4}} c_\sigma^{d_2} M_4^{ (2)} (1, \sigma, 6),\label{eq:d26pe}
\eea
where the coefficients $c_\sigma^{d_2}$ corresponding to the 24 orderings of $\sigma$ are given in Table \ref{tab:6ptcoe}. Just like the $4$-pt case in Eq. (\ref{eq:coes4p}), the above is the unique solution modulo ${\cal O}(p^2)$ BCJ relations among $M_n^{ (2)} (1, \sigma, n)$, which in general have the form
\bea
\sum_{\sigma \in S_{n-2}} c_\sigma^{(\text{BCJ})} M_n^{ (2)} (1, \sigma, n) = 0\ ,
\eea
with $c_\sigma^{(\text{BCJ})} = \ordr (p^2)$ being local. For example, the ``fundamental BCJ relation'' is \cite{Bern:2008qj,Feng:2010my}:
\bea
s_{12}M_n^{ (2)} (\mathbb{I}_n)+ \sum_{i=3}^{n-1} \left( \sum_{j=1}^{i} s_{2j}\right) M_n^{ (2)} (1, 3,4,\cdots,i,2,i+1,\cdots,n-1, n)=0.
\eea
There are $(n-2)!-(n-3)!$ independent relations.
\begin{table}
\centering
\begin{tabular}{c|c}
\hline
$\sigma$ & $(4 \Lambda^2 ) c_\sigma^{d_2}$\\
\hline
$2345$ & $2(s_{23} - s_{13})$\\
$2354$, $2534$, $3254$, $3524$, $5234$, $5324$ & $0$\\
$2435$, $2453$ & $2(s_{14} - s_{24})$\\
$2543$ & $2(s_{25} - s_{15})$\\
$3245$ & $-2 s_{13}$\\
$3425$ & $s_{56} - 2s_{13}$\\
$3452$ & $s_{14} + s_{34} - s_{26}$\\
$3542$, $5342$ & $-P^2_{135} $\\
$4235$, $4253$ & $2s_{14}$ \\
$4325$ & $2s_{14} - s_{56}$\\
$4352$, $4532$ & $ s_{14} + s_{26}$\\
$4523$ & $ 2s_{14} - s_{36} $\\
$5243$ & $-2 s_{15}$\\
$5423$ & $ s_{35} - 2s_{15}$\\
$5432$ & $ s_{23} + s_{36} - s_{15} $\\
\hline
\end{tabular}
\caption{The solutions for $c_\sigma^{d_2}$, which has a common factor of $1/(4 \Lambda^2 )$.}\label{tab:6ptcoe}
\end{table}

\begin{figure}[htbp]
\captionsetup[subfigure]{position=b}
\centering
\subcaptionbox{The half-ladder graph $g_1$\label{fig:6hl}}{\includegraphics[width=.59\linewidth]{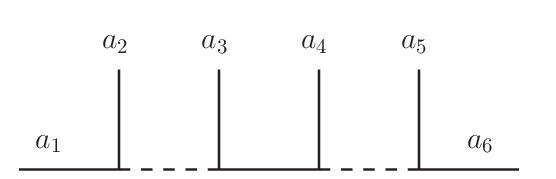}}
\subcaptionbox{The ``star'' graph $g_2$\label{fig:6st}}{\includegraphics[width=.4\linewidth]{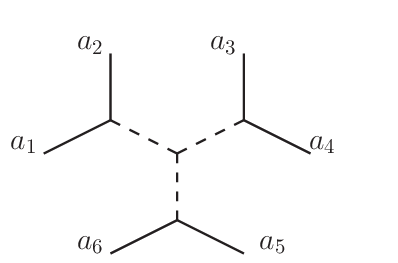}}
\caption{Two kinds of cubic graphs at $6$-pt.}
\label{fig:6pcg}
\end{figure}

The full amplitude corresponding to $M_{d_2, 6}^{ (4)}$ is
\bea
{\cal M}_{d_2, 6}^{ (4)} = \sum_{i = 1}^5 \sum_{j=i+1}^6 \sum_{\sigma \in S_4/Z_4} \delta^{a_i a_j} \tr \left[ X^{a_{\sigma (1)}} X^{a_{\sigma (2)}} X^{a_{\sigma (3)}} X^{a_{\sigma (4)}} \right] M_{d_2, 6}^{ (4)}  (i,j|\sigma),
\eea
where $\sigma$ are permutations of $\{ 1,2, \cdots, 6 \} \setminus \{i,j \}$ modulo cyclic permutations. From Eq. (\ref{eq:d26pe}) we know that the above can be expanded in the form of Eq. (\ref{eq:npd}), where the modified flavor numerators $\hat{\sff}_{d_2,g_1}$ for the half-ladder graph given in Fig. \ref{fig:6hl} has the following relabeling symmetric form:
\begin{align}
\nonumber&\hat{\sff}_{d_2,g_1 }= \frac{1}{4\Lambda^2 }\times \\
\nonumber&\Big\{(s_{13}+s_{23}) \left[2 T^i_{a_1 a_2} (T^i_{a_3 a_6}-T^i_{a_4 a_5}+T^i_{a_4 a_6})-T^i_{a_5 a_6} (T^i_{a_1 a_3}-T^i_{a_2 a_3}+2 T^i_{a_3  a_4})\right.\\
\nonumber &\qquad \qquad \quad \left.+T^i_{a_3 a_5} (T^i_{a_2  a_4}-T^i_{a_1  a_4})-T^i_{a_3 a_6} (T^i_{a_2  a_4}-T^i_{a_1  a_4})\right]\\
\nonumber&+s_{12} \left[T^i_{a_1 a_2} (2 T^i_{a_3 a_6}-T^i_{a_4 a_5}+T^i_{a_4 a_6}-2 T^i_{a_5 a_6})+T^i_{a_5 a_6} (T^i_{a_1 a_3}+2 T^i_{a_1  a_4}-T^i_{a_2 a_3}-2 T^i_{a_2  a_4})\right.\\
\nonumber &\qquad \quad +T^i_{a_1 a_3} (T^i_{a_4 a_6}- T^i_{a_4 a_5})+T^i_{a_1  a_4} (T^i_{a_3 a_6} -  T^i_{a_3 a_5})+T^i_{a_2 a_3}( T^i_{a_4 a_5}- T^i_{a_4 a_6})\\
\nonumber&\left.+T^i_{a_2  a_4}( T^i_{a_3 a_5}- T^i_{a_3 a_6})\right] -(s_{45}+ s_{46}) \left[2 T^i_{a_1 a_2} (T^i_{a_3 a_6}+T^i_{a_4 a_6}-T^i_{a_5 a_6}  T^i_{a_4 a_5} ) \right.\\
\nonumber &\qquad \qquad \qquad \qquad \quad \left.-T^i_{a_1 a_3} (T^i_{a_5 a_6}-T^i_{a_4 a_6} + T^i_{a_4 a_5})+T^i_{a_2 a_3} (T^i_{a_4 a_5}-T^i_{a_4 a_6}+T^i_{a_5 a_6})\right]\\
&-2 T^i_{a_1 a_2} T^i_{a_3  a_4} s_{46} -2 T^i_{a_1 a_2} T^i_{a_3 a_5} s_{56} \Big\}\, ,
\end{align}
where $T^i_{a_k a_l} \equiv (T^i)_{a_k a_l}$ is the matrix entry of the group generator $T^i$ in the anti-symmetric basis. Note that we have omitted factors of $\delta^{a_i a_j}$, which can easily be restored from the two missing flavor labels in each term. The modified flavor factor of the other kind of 6-pt cubic graph, as shown in Fig. \ref{fig:6st}, can then be directly calculated using the Jacobi relations.

We checked that the two kinds of 8-pt partial amplitudes $M_{d_2, 8}^{ (4)} (1,2|3,4,5,6,7,8) $ and $M_{d_2, 8}^{ (4)} (1,2,3,4|5,6,7,8)$, computed in \cite{Bijnens:2019eze} and given by $\nlsm^{d_2}$ also admit expansions in the form of Eq. (\ref{eq:exp}), with local coefficients $c_\sigma^{(d_2)}$. In other words, ${\cal M}_{d_2, 8}^{ (4)}$ also can be expanded as in Eq. (\ref{eq:npd}) with modified flavor factors $\hat{\sff}_{d_2,g }$ that are local.  To obtain them, one has to use an ansatz of the form
\begin{align}
M_8^{(4)}=\sum_{i, \sigma \in S_{6}} x_{i,\sigma}\, s_{i} \, M_8^{(2)}(1,\sigma,8)
\end{align}
where $\{s_i\}=\{s_{12},s_{13},\ldots\}$ is the list of 20 independent Mandelstam invariants at 8-pt. This ansatz, which contains 14400 unfixed parameters $x_{i,\sigma}$, requires significant computation time to solve. Once this expansion is established, finding the corresponding modified flavor numerators $\hat{\sff}_{d_2,g}$ is straightforward. The resulting expressions are long and not particularly illuminating. They will not be presented here but can be available upon request to the authors.

\section{The double copy relations}

\label{sec:dc}

Given the flavor-kinematics duality demonstrated for a particular linear combination of ${\cal O}(p^4)$ operator, it is natural to ask if there is a double copy relation. At $\ordr (p^2)$, the $n$-pt $\nlsm^{(2)}$ amplitudes can be written as the following
\bea
{\cal M}_{n}^{(2)} = \frac{1}{f^{n-2}} \sum_{g \in \{g_n\} } \frac{ \sff_{R,g}\ n_{g}^\nl}{d_{g}},\label{eq:cknl2}
\eea
where $\sff_{R,g}$ are flavor factors corresponding to cubic graph $g$ and expressed in terms of generators in the representation $R$, the 4-pt example of which is given in Eq. (\ref{eq:frdef}). For the specific case of $R$ to be the adjoint representation $A$, $\sff_{A,g} $ is the same as the color factor in the YM amplitude $\mathsf{c}_g$, as in Eq. (\ref{eq:ckym}). Replacing the $\nlsm^{(2)}$ kinematic numerator $n_{k,g}^\nl$ in Eq. (\ref{eq:cknl2}) with another copy of color factor $\tilde{\mathsf{c}}_g$ (of a different group), we arrive at (up to coupling constants)
\bea
{\cal M}_{n}^{\phi^3} = \sum_{g \in \{g_n\} } \frac{ \mathsf{c}_g\ \tilde{\mathsf{c}}_g }{d_{g}},\label{eq:ckbiadj}
\eea
which is the tree amplitude for the cubic bi-adjoint scalar theory $\phi^3$, generated by the Lagrangian
\bea
\label{eq:cubicbiL}
\lag_{\phi^3} = \frac{1}{2} \partial_\mu \phi^{a \ta} \partial^\mu \phi^{a \ta} - \frac{\lambda}{6} \phi^{a \ta} \phi^{b \tb} \phi^{c \tc} f^{abc} \tdf^{\ta \tb \tc},
\eea
where each scalar field $\phi^{a \ta}$ carries two labels, $a$ for the adjoint of group $G$ and $\ta$ is for the adjoint of group $\tilde{G}$; $f^{abc}$ and $\tdf^{\ta \tb \tc}$ are the structure constants for $G$ and $\tilde{G}$, respectively. It is well understood that there is an intimate connection between the double copy structure and the KLT relations \cite{Bern:2019prr}, and  Eqs.~(\ref{eq:cknl2}) and (\ref{eq:ckbiadj}) leads to the (trivial) KLT relation for $\nlsm^{(2)}$ \cite{Cachazo:2013iea,Cachazo:2014xea}:
\bea
M^{(2)} (\mathbb{I}_n) = \sum_{\alpha, \beta}  M^{(2)} (1,\alpha,n-1,n) \ S_n (\alpha || \beta ) \ M^{\phi^3} (1,\alpha,n,n-1||\mathbb{I}_n) ,\label{eq:nlklt}
\eea
where $M^{\phi^3} (\sigma_1||\sigma_2)$ is the doubly ordered amplitudes for $\phi^3$, and
\bea
S_n (\alpha || \beta) = \left[  M^{\phi^3}_n (1,\alpha,n-1,n||1,\beta,n,n-1)  \right]^{-1}
\eea
is the KLT kernel \cite{Cachazo:2013iea}. Eq.~(\ref{eq:nlklt}) is trivial in the sense that we are multiplying the NLSM$^{(2)}$ amplitudes by unity, as $S_n (\alpha || \beta)$ is the inverse of the cubic bi-adjoint amplitudes. What is less trivial is the fact that there is a universal KLT kernel for theories of adjoint fields, which subsequently is identified with the inverse of the cubic bi-adjoint amplitudes \cite{Cachazo:2013iea}. Notice that in the partial amplitude $M^{\phi^3}$ we use the double line ``$||$'' to separate the orderings of two different  groups, in contrast to the double trace structure of a single  group, where we use a single line ``$|$'' to denote the separation of the traces. Eq. (\ref{eq:nlklt}) can be written more compactly as 
\bea
\nlsm^{(2)} =  \nlsm^{(2)} \stackrel{\rm KLT}{\otimes} \phi^3 \ .\label{eq:p2KLT}
\eea

For the ${\cal O}(p^4)$ operator that exhibits flavor-kinematic duality
\bea
 \M_{d_2,n}^{(4)} = \frac{1}{f^{n-2}} \sum_{g \in \{g_n\} } \frac{\hat{\sff}_{d_2,g}\ n_{g}^\nl}{d_{g}} \ ,
\eea
the natural question to ask is what happens when we replace $n_{g}^\nl$ with $\tilde{\mathsf{c}}_g$ in the above?\footnote{The other possibility of replacing $\hat{\sff}_{d_2,g}$ by $\tilde{\mathsf{c}}_g$ simply gives back the $\nlsm^{(2)}$ amplitudes.} In other words, can the object
\bea
\sum_{g \in \{g_n\} } \frac{\hat{\sff}_{d_2,g}\ \tilde{\mathsf{c}}_g}{d_{g}}\label{eq:scaobj}
\eea
be interpreted as the S-matrix elements of a consistent quantum field theory? 

First of all, such a theory, if one exists, would contain amplitudes where all external states are scalars that transform under two symmetry groups. Secondly, since $\hat{\sff}_{d_2,g} = \ordr (p^2)$, the amplitude is $\ordr (p^{8-2k})$ and has a mass dimension higher by 2 compared to that of $\phi^3$. For example, the 4-pt amplitude from Eq. (\ref{eq:scaobj}) is (up to coupling constants)
\bea
&& - f_{\delta, s}  \tilde{\mathsf{c}}_s-  f_{\delta, t}  \tilde{\mathsf{c}}_t  - f_{\delta, u}  \tilde{\mathsf{c}}_u\nonumber \\
 &&+ \frac{\delta^{a_1 a_2} \delta^{a_3 a_4} (t-u) \tilde{\mathsf{c}}_s }{s} + \frac{\delta^{a_2 a_3} \delta^{a_1 a_4} (u-s) \tilde{\mathsf{c}}_t }{t} + \frac{\delta^{a_1 a_3} \delta^{a_2 a_4} (s-t) \tilde{\mathsf{c}}_u}{u} \ , \label{eq:gyms4}
\eea
which contains poles in the $s$-, $t$- and $u$-channels. This suggests non-vanishing 3-pt vertices in the theory. However, a massless scalar 3-pt vertex can be non-zero on-shell only if it does not carry momentum dependence, which in turn would not give the correct mass dimension $\ordr (p^0)$ in Eq. (\ref{eq:gyms4}). We are thus led to the conclusion that the poles in Eq. (\ref{eq:gyms4}) must come from intermediate vector states!

The $s$-channel residue of Eq. (\ref{eq:gyms4}) is
\bea
\delta^{a_1 a_2} \delta^{a_3 a_4} (t-u) \tilde{\mathsf{c}}_s = \left[i \delta^{a_1 a_2} f^{\tilde{i} \ta_1 \ta_2} (p_1 - p_2)  \right] \cdot \left[i \delta^{a_3 a_4} f^{\tilde{i} \ta_3 \ta_4} (p_3 - p_4)  \right] \label{eq:4ptmass2} ,
\eea
which leads us to deduce that the three point vertex in this theory is
\bea
i \delta^{a_1 a_2} f^{ \ta_1 \ta_2 \ta_3} (p_1 - p_2)^{\mu_3},\label{eq:gyms3p}
\eea
where legs $1$ and $2$ are scalars $\phi^{a\ta}$ carrying two adjoint indices, $a$ of group $G$ and $\ta$ of group $\tilde{G}$, while leg $3$ is the vector boson $A_\mu^\ta$ carrying the adjoint representation of $\tilde{G}$, as well as a Lorentz index $\mu_3$. Such a vertex naturally arises in the following gauged kinetic term of the scalars:
\bea
\frac{1}{2} \tilde{\tr} \left( D_\mu \uphi^a D^\mu \uphi^a \right),
\eea
where $D_\mu = \partial_\mu + igA_\mu^\ta \tilde{\gX}^\ta$ is the gauge covariant derivative, $\tilde{\gX}$ is the generator for $\tilde{G}$, $g$ being the gauge coupling, and $\uphi^a \equiv \phi^{a \ta} \tilde{\gX}^\ta$. Therefore we will call $G$ the flavor group and $\tilde{G}$ the gauge group.

Assuming the 3-pt vertex given in Eq. (\ref{eq:gyms3p}), the 4-pt contact term in Eq. (\ref{eq:gyms4}) can come from the following interaction:
\bea
-f^{\ta_1 \ta_2 \tb} f^{\ta_3 \ta_4 \tb} \phi^{a \ta_1} \phi^{b \ta_2} \phi^{a \ta_3} \phi^{b \ta_4} = \tilde{\tr} \left( [\uphi^a,\uphi^b]^2 \right) \ .
\eea
Including the propagators for the massless vector states, we arrive at the following Lagrangian by examining the 4-pt amplitude in Eq.~(\ref{eq:4ptmass2}),
\bea
\label{eq:sYMSL}
{\cal L}_{\rm YMS}=\tilde{\tr} \left( \frac{1}{4} \uf^{ \mu \nu} \uf_{\mu \nu} + \frac{1}{2} D_\mu \uphi^a D^\mu \uphi^a - \frac{g^2}{4} [\uphi^a,\uphi^b]^2 \right)\ ,
\eea
where $\uf_{ \mu \nu} = F^{\ta}_{\mu \nu} \tilde{\gX}^\ta$ is the field strength tensor of the gauge bosons. This is the Lagrangian for the well-known Yang-Mills scalar (YMS) theory, which can be seen as a dimensional reduction of the YM theory. Consequently,  at the 4-pt level we are led to the observation that the following KLT relation holds (up to coupling constants):
\bea
M_{d_2,4}^{(4)} (1,2|3,4) =  tu =  (u) (s) \left(\frac{t}{s}\right) = M_4^{(2)} (1,2,3,4) S_4 (2||2) M_4^{\rm YMS} (1,2|3,4||1,2,4,3) \ ,\ \ 
\eea
with $M_n^{\rm YMS} (\alpha|\beta||\sigma)$ being the partial amplitudes of YMS that is ordered in both the flavor and color groups.

Beyond the 4-pt level, it is important to recall that the 4-pt ${\cal O}(p^4)$ amplitude $\M_{d_2,4}^{(4)}$ cannot exist on its own in a consistent quantum field theory; it is part of the derivative expansion in the $2\to 2$ scattering amplitude that starts at ${\cal O}(p^2)$,
\bea 
\M_n^{{\rm NLSM}^{d_2}} = {\cal M}_{n}^{(2)}+ \M_{d_2,n}^{(4)} +{\cal O}(p^6) \ ,
\eea
where NLSM$^{d_2}$ is a  quantum field theory containing ${\cal O}(p^2)$ NLSM amplitudes and the ${\cal O}(p^4)$ amplitudes soft-bootstrapped from the soft block ${\cal S}_2^{(4)} (1,2|3,4)$. Recall at ${\cal O}(p^2)$ the amplitudes already have the double copy relation involving the cubic bi-adjoint scalars in Eq.~(\ref{eq:p2KLT}). This suggests a consistent double copy relation must also include the cubic bi-adjoint scalar vertex in Eq.~(\ref{eq:cubicbiL}), in addition to those in Eq.~(\ref{eq:sYMSL}).

It turns out a theory with the Lagrangian ${\cal L}_{\rm YMS}+\lag_{\phi^3}$ has been studied previously and dubbed  the  YM$+\phi^3$ theory \cite{Chiodaroli:2014xia}, also called ``generalized Yang-Mills scalar theory'' in \cite{Cachazo:2014xea}, which is generated by the following Lagrangian:
\bea
\lag_{\textrm{YM}+\phi^3} = \tilde{\tr} \left( \frac{1}{4} \uf^{ \mu \nu} \uf_{\mu \nu} + \frac{1}{2} D_\mu \uphi^a D^\mu \uphi^a - \frac{g^2}{4} [\uphi^a,\uphi^b]^2 \right) - \frac{\lambda}{6} \phi^{a \ta} \phi^{b \tb} \phi^{c \tc} f^{abc} \tdf^{\ta \tb \tc}. \label{eq:Lgyms}
\eea
The above can be seen as a specific ``higher derivative'' extension of the cubic bi-adjoint scalar theory $\phi^3$, where the group $\tilde{G}$ is gauged. So the conjectured double copy structure has the following KLT bilinear relation
\bea
M_{d_2,n}^{(4)} (\alpha|\beta) = \sum_{\sigma_1, \sigma_2}  M^{(2)}_n (1,\sigma_1,n-1,n)\ S_n (\sigma_1 || \sigma_2 ) \ M^{{\textrm{YM}+\phi^3}}_n (\alpha|\beta||1,\sigma_2,n,n-1) ,\label{eq:kltd2}
\eea
where  $M^{{\textrm{YM}+\phi^3}}_n (\alpha|\beta||\sigma)$ is the doubly ordered amplitude for YM+$\phi^3$ with scalar external states. In YM+$\phi^3$, when all external states are scalars, the $n$-pt amplitude at the lowest order in the derivative expansion, which is $\ordr (p^{6-2n})$, coincides with ``pure'' the bi-adjoint scalar theory amplitude $\M^{\phi^3}_n$. On the other hand, YM+$\phi^3$ scalar amplitudes of higher orders in the derivative expansion actually coincide with the $\nlsm \oplus \phi^3$ theory discussed in Refs. \cite{Cachazo:2016njl,Mizera:2018jbh}, when all external scalars are bi-adjoint. Using the 6-pt amplitudes provided in Ref. \cite{Mizera:2018jbh}, we have checked explicitly that the KLT relation in Eq. (\ref{eq:kltd2}) holds at $n=6$.

From the KLT relation in Eq.~(\ref{eq:kltd2}) one may be tempted to conclude $\nlsm^{d_2}$ as a double copy of $\nlsm^{(2)}$ and ${\textrm{YM}+\phi^3}$, 
\bea
\nlsm^{d_2} \stackrel{?}{=}  \nlsm^{(2)} \stackrel{\rm KLT}{\otimes} \left({\textrm{YM}+\phi^3}\right).\label{eq:dcd2}
\eea
However, this is clearly not true as ${\textrm{YM}+\phi^3}$ is a theory which also contains gluons as external particles. There are also trace and flavor structures in ${\textrm{YM}+\phi^3}$ that are not present in the NLSM and Eq.~(\ref{eq:scaobj}) only covers a subset of scalar amplitudes in ${\textrm{YM}+\phi^3}$. In Eq. (\ref{eq:scaobj}) each flavor trace always involves an even number of generators. However, the double-trace components in $\M^{\textrm{YM}+\phi^3}$ also include non-vanishing terms where each flavor trace contains an odd number of generators. The theory of $\nlsm^{d_2}$ does not generate modified flavor factors $\hat{\sff}_{d_2,g}$ of these kinds, but ${\textrm{YM}+\phi^3}$ does. Therefore, the double copy relation of  Eq. (\ref{eq:dcd2}) holds only in the limited sense of relations between partial amplitudes as in Eq. (\ref{eq:kltd2}), where the orderings $\alpha$ and $\beta$ both contain an even number of labels.

So, more precisely, the double copy relation should read
\bea
\nlsm^{d_2} \subset  \nlsm^{(2)} \stackrel{\rm KLT}{\otimes} \left({\textrm{YM}+\phi^3}\right).\label{eq:dcd4}
\eea
What is the theory that is the double copy of $\nlsm^{(2)}$ and ${\textrm{YM}+\phi^3}$ then? It turns out the question has been studied using the CHY formalism in Ref.~\cite{Cachazo:2014xea} and the theory is an  EFT called the extended Dirac-Born-Infeld (eDBI) theory \cite{Cachazo:2014xea}, which involves scalars $\pi^a$ with flavors, and also a  $\U (1)$ gauge boson $A_\mu$. The Lagrangian of eDBI is given by
\bea
\lag_{\edbi} =- \frac{f^2 \Lambda^2}{2} \left[ \sqrt{ - \det \left( \eta_{\mu \nu}  - \frac{2}{\Lambda^2} \left(    
\< d_\mu d_\nu \> + F_{\mu \nu} + W_{\mu \nu} \right) \right) }- 1\right],
\eea
where $F_{\mu \nu} = \partial_\mu A_\nu - \partial_\nu A_\mu$, and $|d_\mu\>$ is the same as that in the NLSM defined in Eq. (\ref{eq:nlddef}). On the other hand, $W_{\mu \nu}$ is an infinite sum of terms, each being a single trace of odd powers of $\pi^a \gX^a/f$ with two derivatives $\partial_\mu$ and $\partial_\nu$ acting on it; $W_{\mu \nu}$ is also anti-symmetric in $\mu$ and $\nu$.\footnote{The exact form of $W_{\mu \nu}$ is not relevant for this work. Ref. \cite{Cachazo:2014xea} contains an expression for $W_{\mu \nu}$ in the Cayley parameterization \cite{Kampf:2013vha} of the scalars; this cannot be used here as we are working in the exponential parameterization as shown in Eq. (\ref{eq:lagst}). To our knowledge $W_{\mu \nu}$ in the exponential representation has not been written down.}  It has been proved  using the CHY formalism that
\bea
\edbi = \nlsm^{(2)} \stackrel{\rm KLT}{\otimes} \left({\textrm{YM}+\phi^3}\right).
\eea
Brief overviews of the CHY formalism and the above double copy relation are  provided in appendices.

For our purpose, it is instructive to expand $\lag_{\edbi}$ to $\ordr (p^4)$, 
\bea
\lag_{\edbi} = \lag_\nlsm^{d_2} + \frac{f^2}{2 \Lambda^2} \left( F_{\mu \nu} + W_{\mu \nu} \right)^2+  \ordr \left(\frac{1}{\Lambda^4} \right),
\eea
where $\lag_\nlsm^{d_2}$ is exactly the Lagrangian of $\nlsm^{d_2}$:
\bea
\lag_\nlsm^{d_2} = \lag_\nlsm^{(2)} + \frac{f^2}{\Lambda^2} \left(-\frac{1}{4} O_1 + \frac{1}{2} O_2 \right).\label{eq:lagd2}
\eea
At $\ordr (p^4)$, the difference between ${\cal M}_{\edbi, n}^{ (4)}$ where all external particles are scalars, and ${\cal M}_{d_2, n}^{ (4)}$, is that the former also contains double trace flavor structures where each trace contains odd powers of $\gX^a$: these are contributions of the term $( F_{\mu \nu} + W_{\mu \nu} )^2$, and there can exist internal photons in the Feynman diagrams of these amplitudes. They are naturally generated in the KLT relation from ${\textrm{YM}+\phi^3}$, but are absent in $\nlsm^{d_2}$. However, if we restrict to the partial amplitudes with an even number of NGB's in each trace, then the contributions of $\lag_\nlsm^{d_2}$ and $\lag_{\edbi}$ are exactly the same. Therefore, we can write down the CHY formulas for these amplitudes, as the CHY representation of eDBI is known \cite{Cachazo:2014xea}:
\bea
M_{d_2, n}^{ (4)} (\alpha|\beta) = \oint d \mu_n\ \pt (\alpha) \pt (\beta) \Pf' \Pi_n (\alpha \cup \beta) (\Pf' \A_n )^2,\label{eq:chyd2}
\eea
up to coupling constants.

\section{Conclusion and Discussions}\label{sec:con}
In this work we have explored the possibility of extending the flavor-kinematics duality to $\mathcal{O}(p^4)$ operators in NLSM, by using new modified flavor numerators that mix flavor and kinematic factors. While at 4-pt all four operators have such a flavor-dual representation, we find that at 6-pt this is true only for one particular operator, corresponding to double-trace amplitudes. Furthermore, these specific amplitudes are seen to coincide with a subset of amplitudes given by eDBI, which is the double copy of $\nlsm^{(2)}$ and ${\textrm{YM}+\phi^3}$.

It remains an open question whether there are further modifications that must be made to numerators in order to accommodate the remaining operators at $\mathcal{O}(p^4)$. We have assumed that the modified flavor factors are local; such a constraint may be relaxed in a well-defined way. For example, what happens if we do not require the flavor factors to be local, but still satisfy relabeling symmetry? Also, these non-local flavor factors should not lead to physical amplitudes in the (reversed) double-copy procedure discussed in Section \ref{sec:dc}, though the end products of this procedure may still have a CHY representation.  In known cases it seems there is an intimate connection between theories having a CHY representation and having a double copy relation. Then, are there CHY representations the $\ordr (p^4)$ NLSM amplitudes of other operators?

In principle, scattering amplitudes may be expressed as linear combinations between different pairs of numerators. For instance, the most general possibility is given by: 
\bea
M_6^{(4)}=\sum_{g\in\{g_n\!\}}  \frac{\hat{f}_g^{(0)}n_g^{(10)}+\hat{f}_g^{(2)}n_g^{(8)}+\ldots+\hat{f}_g^{(8)}n_g^{(2)}}{d_g}
\eea
where $\hat{f}_g^{(k)}$ and $n_g^{(k)}$ are modified flavor/color and kinematic numerators of mass dimension $k$ (note that $d_g$ has mass dimension 6 at 6-pt). From a bootstrap perspective, such an ansatz is allowed, but would be very difficult to solve already even at 6-pt for $\mathcal{O}(p^4)$.  A related question is then whether partial amplitudes constructed via modified color/flavor factors satisfy BCJ-like relations, as these are typically much simpler to solve and would indicate the existence of a color-kinematic duality. It would also be interesting to understand whether all trace structures present in ${\textrm{YM}+\phi^3}$ amplitudes have color-dual representations.

Also notice that in our work we have always assumed the ``cubic adjoint'' properties for the flavor structures: they correspond to cubic graphs that satisfy anti-symmetry and Jacobi relations. This is true when all interactions are dressed with structure constants $f^{abc}$, as in the YM theory. For $\nlsm^{(2)}$, this is guaranteed by the closure condition Eq. (\ref{eq:clocon}) for the generators $T^i_{ab}$, while more general flavor structures appear at $\ordr (p^4)$. It is then natural to ask whether some version of flavor-kinematics duality can exist for these more general structures, perhaps involving theories beyond the cubic graphs.

Finally, these considerations can be explored beyond the $\mathcal{O}(p^4)$ operators investigated in this work. Composition rules such as those used in Eqs.~(\ref{eq:defnnl}) that generate color-kinematic solutions can be extended to higher multiplicity \cite{Carrasco5point}, and it would be fascinating to see how they can be used to bootstrap the infinite tower of corrections to the NLSM.

\begin{acknowledgments}
The authors would like to thank Mattias Sj\"{o} for providing expressions for high-point NLSM amplitudes, John Joseph M. Carrasco and Suna Zekioglu for enlightening discussions, as well as Henrik Johansson for helpful comments on the manuscript. IL and ZY would like to thank Mariana Carrillo González, Callum R.T. Jones, Shruti Paranjape, and Mark Trodden for useful discussions. This work is supported in part by the U.S. Department of Energy under contracts No. DE-AC02-06CH11357 and No. DE-SC0010143. L Rodina is supported by the European Research Council under ERC-STG-639729, Strategic Predictions for Quantum Field Theories .

\end{acknowledgments}

\begin{appendix}

\section{Review of the CHY representation}
\label{app:chy}

The formalism proposed by Cachazo, He and Yuan \cite{Cachazo:2013gna,Cachazo:2013hca,Cachazo:2013iea,Cachazo:2014xea} can express the tree amplitudes of many theories in an integral of the following form:
\bea
\M_n  = \oint d\mu_n \; \mathcal{I}_L (\{p, \vep, \sigma \}) \; \mathcal{I}_R (\{p, \tilde{\vep}, \sigma\}),\label{eq:chyg}
\eea
where $\{p \}$ are the on-shell external momenta, $ \{ \vep \}$ are the polarization vectors, $\{ \sigma \} = \{ \sigma_1 , \sigma_2, \cdots, \sigma_n \}$ are $n$ dimensionless complex variables. The integral in the above are evaluated as residues at the poles of $\{ \sigma \}$ given by the scattering equations
\bea
E_j \equiv \sum_{i \ne j} \frac{p_i \cdot p_j}{ \sigma_{ij} }=0
\eea
where $\sigma_{ij}  = \sigma_i - \sigma_j$. This is realized by the measure
\bea
d \mu_n &\equiv& ( \sigma_{ij}\sigma_{jk}\sigma_{ki}) (\sigma_{pq}\sigma_{qr}\sigma_{rp}) \prod_{a \neq i,j,k} E_a^{-1} \prod_{b \neq p,q,r} d\sigma_b.
\eea
The choice of $\{i, j, k\}$ does not affect the value of the integral, which can be treated as some ``gauge invariance.''

\subsection{The CHY integrands for relevant theories}

Now let us discuss different integrands $\mathcal{I}_{L/R}$ for the theories relevant in our paper. The CHY formula for the doubly ordered partial amplitudes for the cubic bi-adjoint scalar theory $\phi^3$ is 
\bea
M^{\phi^3}_n (\alpha || \beta) = \oint d\mu_n ~\mathcal{C}_n  (\alpha)~\mathcal{C}_n  (\beta),
\eea
where $\mathcal{C}_n$ is the Parke-Taylor factor:
\bea
\mathcal{C}_n (\mathbb{I}_n) = \frac{1}{\sigma_{12} \sigma_{23} \cdots \sigma_{n-1,n} \sigma_{n,1}}.
\eea
Notice that $\mathcal{C}_n$ is not permutation invariant and contains the ordering for the trace basis. The flavor-ordered amplitudes for $\nlsm^{(2)}$ is
\bea
M_n^{(2)} (\alpha) =  \oint d\mu_n ~(\pf' \A_n)^2 ~\mathcal{C}_n  (\alpha),
\eea
where the anti-symmetric matrix $\A_n$ is given by
\bea
[\A_n]_{ab} =\left\{ \begin{array}{ll}\dfrac{ p_a \cdot p_b}{\sigma_{ab}}, & a \neq b, \\	0, & a = b.
	\end{array} \right.,\label{eq:chyma}
\eea
and the reduced Pfaffian $\pf'$ is defined as $\pf' \A_n = \frac{(-)^{a+b}}{\sigma_{ab}} \pf \A_n^{[a,b]}$, with $\A_n^{[a,b]}$ being the matrix $\A_n$ with rows and columns of labels $a$ and $b$ removed. Again, the choice of $a$ and $b$ does not affect the value of the CHY integral.

The eDBI is a theory involving scalars and photons, the flavor-ordered amplitudes of which are given by
\bea
M_n^{\edbi} (\alpha_1 | \alpha_2 | \cdots | \alpha_m ) = \oint d \mu_n\ \left[ \prod_{i=1}^m \mathcal{C} (\alpha_i) \right] \Pf' \Pi_n (\alpha_1 \cup \alpha_2 \cup \cdots \cup \alpha_m) (\Pf' \A)^2.\label{eq:chyedbi}
\eea
The above is an $n$-pt amplitude with $m$ traces, and $\alpha_1 \cup \alpha_2 \cup \cdots \cup \alpha_m$ contain labels for the scalars, as they carry flavors and admit orderings; the rest of the labels, in $V = \{1,2, \cdots , n\} \setminus ( \alpha_1 \cup \alpha_2 \cup \cdots \cup \alpha_m)$, are for the photons. In the following, we will denote the size of $V$ (i.e. the number of external photons) as $v$, use $a,\ b$ to denote labels in $V$, and $i, \ j$ to denote labels for the traces, i.e. $i,\ j = 1,2,\cdots, m$. The $2(m+v) \times 2(m+v)$ anti-symmetric matrix $\Pi$ in Eq. (\ref{eq:chyedbi}) is given by
\bea
\Pi_n (\alpha_1 \cup \alpha_2 \cup \cdots \cup \alpha_m) =\left( \begin{array}{cccc} \A & - \left[ \textsf{P}^{(1)} \right]^T  & - \textsf{C}^T & - \left[ \textsf{P}^{(4)} \right]^T \\	\textsf{P}^{(1)} & \textsf{P}^{(2)} &   \textsf{P}^{(3)}  & - \left[ \textsf{P}^{(5)} \right]^T\\ \textsf{C} & - \left[ \textsf{P}^{(3)} \right]^T & \textsf{B} & - \left[ \textsf{P}^{(6)} \right]^T \\ \textsf{P}^{(4)} & \textsf{P}^{(5)} & \textsf{P}^{(6)} & \textsf{P}^{(7)}
	\end{array} \right),\label{eq:chydpi}
\eea
where $\A$, $\textsf{B}$ and $\textsf{C}$ are $v\times v$ square matrices determined by the information of the photons, with $\A$ defined in Eq. (\ref{eq:chyma}), and
\bea
[\textsf{B}]_{ab} =\left\{ \begin{array}{ll}\dfrac{ \vep_a \cdot \vep_b}{\sigma_{ab}}, & a \neq b, \\	0, & a = b.
	\end{array} \right., \qquad [\textsf{C}]_{ab} =\left\{ \begin{array}{ll}\dfrac{ \vep_a \cdot p_b}{\sigma_{ab}}, & a \neq b, \\	\displaystyle-\sum\limits_{c \in V ,\ c \ne a} \dfrac{2 \vep_a \cdot p_c}{\sigma_{ab}}, & a = b.
	\end{array} \right.;
\eea
$ \textsf{P}^{(1,3,4,6)}$ are $v \times m$ matrices that have both photon and trace indices:
\bea
\begin{array}{ll}
\displaystyle\left[ \textsf{P}^{(1)} \right]_{ia} =\sum_{r \in \alpha_i} \frac{ p_r \cdot p_a}{\sigma_{ra}},& \displaystyle \left[ \textsf{P}^{(3)} \right]_{ia} =\sum_{r \in \alpha_i} \frac{ p_r \cdot \vep_a}{\sigma_{ra}}, \\ \displaystyle\left[ \textsf{P}^{(4)} \right]_{ia} =\sum_{r \in \alpha_i} \frac{ \sigma_r p_r \cdot p_a}{\sigma_{ra}}, &\displaystyle \left[ \textsf{P}^{(6)} \right]_{ia} =\sum_{r \in \alpha_i} \frac{\sigma_r p_r \cdot \vep_a}{\sigma_{ra}};\end{array}
\eea 
$ \textsf{P}^{(2,5,7)}$ are $m \times m$ matrices of trace indices:
\bea
&&\left[ \textsf{P}^{(2)} \right]_{ij} =\sum_{r_1 \in \alpha_i,r_2 \in \alpha_j} \frac{ p_{r_1} \cdot p_{r_2}}{\sigma_{r_1 r_2}}, \qquad \left[ \textsf{P}^{(5)} \right]_{ij} =\sum_{r_1 \in \alpha_i,r_2 \in \alpha_j} \frac{ \sigma_{r_1} p_{r_1} \cdot p_{r_2}}{\sigma_{r_1 r_2}},\\
&& \left[ \textsf{P}^{(7)} \right]_{ij} =\sum_{r_1 \in \alpha_i,r_2 \in \alpha_j} \frac{ \sigma_{r_1} \sigma_{ r_2}p_{r_1} \cdot p_{r_2}}{\sigma_{r_1 r_2}}.
\eea
It is understood in the above that $r_1 \ne r_2 $ is applied in the sum for the diagonal elements of these matrices. The reduced Pfaffian $\pf'$ for $\Pi$ is defined as $\pf' \Pi_n =  \pf \Pi_n^{[i,j]}$, where the row $i$  belongs to those rows involving $\textsf{P}^{(2)}$ in Eq. (\ref{eq:chydpi}), the column $j$  belongs to those involving $\textsf{P}^{(7)}$, and they are deleted from $\Pi$ in the reduced Pfaffian.

For the special case when $m=2$ and $v=0$, i.e. the double trace amplitudes of no external photons, the matrix $\Pi$ in Eq. (\ref{eq:chyedbi}) is reduced to the following $4 \times 4 $ matrix:
\bea
\Pi_n (\alpha \cup \beta) =\left( \begin{array}{cc} 	 \textsf{P}^{(2)} &   - \left[ \textsf{P}^{(5)} \right]^T\\  \textsf{P}^{(5)} & \textsf{P}^{(7)}
	\end{array} \right).
\eea
This is what appears in the CHY formula for $\nlsm^{d_2}$.

The YM+$\phi^3$  theory involves scalars of both colors and flavors, as well as gauge bosons with colors. In the following amplitude, the left orderings are for the flavor indices and the right orderings are for the color indices:
\bea
M^{{\textrm{YM}+\phi^3}}_n (\alpha_1 | \alpha_2 | \cdots | \alpha_m ||\beta) = \oint d \mu_n \ \left[ \prod_{i=1}^m \mathcal{C} (\alpha_i) \right]  \Pf' \Pi_n (\alpha_1 \cup \alpha_2 \cup \cdots \cup \alpha_m) \mathcal{C} (\beta),
\eea
where $\alpha_1 \cup \alpha_2 \cup \cdots \cup \alpha_m$ contains labels for the bi-index scalars, and the rest are the gauge bosons. When there are no external gauge bosons, the above formula coincides with the theory of $\nlsm \oplus \phi^3$ \cite{Cachazo:2016njl,Mizera:2018jbh} when all external states are bi-adjoint scalars.

\subsection{KLT relations}

In the CHY representation, if an amplitude of theory ${\cal X}$ is given by
\bea
\M_n^{\cal X} = \oint d \mu_n \ \cin_L \cin_R,
\eea
we can rewrite it as
\bea
\M_n^{\cal X} = \sum_{\alpha, \beta}  \left( \oint d \mu_n \ \cin_L \pt (1,\alpha,n-1,n)  \right) S_n (\alpha || \beta ) \left( \oint d \mu_n' \ \cin_R \pt (1,\alpha,n,n-1)  \right),\label{eq:chyklt}
\eea
where
\bea
S_n (\alpha || \beta) = \left[  M^{\phi^3}_n (1,\alpha,n-1,n||1,\beta,n,n-1)  \right]^{-1}.
\eea
If we can identify in Eq. (\ref{eq:chyklt}) that
\bea
M^{{\cal X}_{L/R}}_n (\alpha) =  \oint d \mu_n \ \cin_{L/R} \pt (\alpha) 
\eea
where $M^{{\cal X}_{L/R}} (\alpha)$ are the ordered amplitudes of some theories ${\cal X}_{L/R}$, then we have the KLT relation:
\bea
\M_n^{\cal X} = \sum_{\alpha, \beta}  M^{{\cal X}_{L}} (1,\alpha,n-1,n) S_n (\alpha || \beta ) M^{{\cal X}_{R}} (1,\alpha,n,n-1) .
\eea
In other words, the theory ${\cal X}$ is a KLT double copy of ${\cal X}_L$ and ${\cal X}_R$: ${\cal X} = {\cal X}_L \stackrel{\rm KLT}{\otimes} {\cal X}_R$.

In the two examples involved in this paper, for
\bea
\nlsm^{(2)} =  \nlsm^{(2)} \stackrel{\rm KLT}{\otimes} \phi^3
\eea
we have
\bea
\cin_L = (\pf' \A_n)^2, \qquad \cin_R = \mathcal{C}_n  (\alpha);
\eea
for
\bea
\edbi = \nlsm^{(2)} \stackrel{\rm KLT}{\otimes} \left({\textrm{YM}+\phi^3}\right)
\eea
we have
\bea
\cin_L = (\pf' \A_n)^2, \qquad \cin_R =\left[ \prod_{i=1}^m \mathcal{C} (\alpha_i) \right] \Pf' \Pi_n (\alpha_1 \cup \alpha_2 \cup \cdots \cup \alpha_m).\label{eq:ilire}
\eea
When we restrict the ordering structure of $\cin_R$ Eq. (\ref{eq:ilire}) to $m=2$, with both $\alpha_1$ and $\alpha_2$ containing an even number of labels, the double copy relation is reduced to
\bea
\nlsm^{d_2} = \nlsm^{(2)} \stackrel{\rm KLT}{\otimes} \left({\textrm{YM}+\phi^3}\right)
\eea

\end{appendix}

\bibliography{references_amp}

\end{document}